\documentclass[11pt]{article}

\def\bb{\bibitem}
\def\lb{\label}
\def\be{\begin{equation}}
\def\ee{\end{equation}}
\def\ba{\begin{eqnarray}}
\def\ea{\end{eqnarray}}

\begin{document}

\title{Comment on `An extreme critical space-time: echoing and
black-hole perturbations'}
\author{G\'erard Cl\'ement
\thanks{Email: gclement@lapp.in2p3.fr}
\\{\small Laboratoire de  Physique Th\'eorique LAPTH (CNRS),} \\
{\small B.P.110, F-74941 Annecy-le-Vieux cedex, France}
\\ \\Sean A. Hayward
\thanks{Email: hayward@mm.ewha.ac.kr}
\\{\small Department of Science Education, Ewha Womans University,} \\
{\small Seodaemun-gu, Seoul 120-750, Korea}}
\date{8 August 2001}
\maketitle

\begin{abstract}
We show that the black hole perturbations of the Hayward static solution to the
massless Einstein-Klein-Gordon equations  are actually gauge artifacts
resulting from the linearization of a coordinate transformation.
\end{abstract}

In a recent paper \cite{hayward}, Hayward discussed a homothetic, static
solution to the massless Einstein-Klein-Gordon equations. With the spherically
symmetric ansatz
\begin{equation}\lb{an}
ds^2=r^2d\Omega^2-2e^{2\gamma}dx^+dx^-
\end{equation}
(where $d\Omega^2$ is the line-element of the unit sphere and $(r,\gamma)$ are
functions of the null coordinates $x^\pm$), this very simple solution reads
\begin{equation}\lb{hay1}
r = (-x^+x^-)^{1/2}\qquad\gamma=0\qquad\phi=\frac12\ln(-x^+/x^-)
\end{equation}
(where $\phi$ is the scalar field). Another form of the solution (\ref{hay1}),
showing explicitly its static and homothetic character, is
\begin{equation}\lb{hay2}
ds^2=e^{2\rho}(d\Omega^2+2d\rho^2-2d\tau^2).
\end{equation}
The relation between the two coordinate systems is
\begin{equation}
x^\pm=\pm e^{\rho\pm\tau}.
\end{equation}

Hayward argued that this solution is critical, in the sense that it lies at the
threshold between black holes and naked singularities. This argument was partly
based on the analysis of the linear perturbations of the static solution
(\ref{hay2}). For a given mode with complex frequency $k$, these linear
perturbations are of the form
\begin{eqnarray}
&&r=e^\rho(1+\epsilon\tilde r(\tau)e^{-k\rho})\lb{pertr}\\
&&\gamma=\epsilon\tilde\gamma(\tau)e^{-k\rho}\\
&&\phi=\tau+\epsilon\tilde\phi(\tau)e^{-k\rho},
\end{eqnarray}
where $\epsilon$ is the perturbation parameter. The resulting linearized field
equations reduce to a single fourth-order ordinary differential equation.
Imposing weak boundary conditions, Hayward found the general solution
\begin{equation}\lb{tr}
\tilde r=A_+e^{\omega_A\tau}+A_-e^{-\omega_A\tau}
+B_+e^{\omega_B\tau}+B_-e^{-\omega_B\tau}
\end{equation}
depending on four integration constants $A_\pm$, $B_\pm$, with the exponents
\begin{eqnarray}
&\omega_A=i\sqrt{(\Im k)^2-3} & \qquad (\Re k = 1)\\
&\omega_B=k & \qquad (\Re k \le 1/2).
\end{eqnarray}

However, as we now show, all the $\omega_B$ modes are spurious, and may be
generated from the static solution (\ref{hay1}) by the coordinate
transformation \ba
x^+ & = & \hat{x}^+ + 2\epsilon B_-(\hat{x}^+)^{1-k} \lb{gauge+}\\
-x^- & = & -\hat{x}^- + 2\epsilon B_+(-\hat{x}^-)^{1-k}\lb{gauge-}, \ea
preserving the double-null ansatz (\ref{an}). To first order in $\epsilon$,
this leads to \be\lb{lin} r = (-x^+x^-)^{1/2} = (-\hat{x}^+\hat{x}^-)^{1/2}(1 +
\epsilon B_-(\hat{x}^+)^{-k} + \epsilon B_+(-\hat{x}^-)^{-k}). \ee Dropping the
hats, this is of the form (\ref{pertr}), with \be \tilde r
=B_+e^{k\tau}+B_-e^{-k\tau}, \ee corresponding to the $\omega_B$ modes of
(\ref{tr}). The perturbative black holes found in \cite{hayward} are therefore
simply artifacts resulting from the linearization of a coordinate
transformation. To see how this happens, note from (\ref{hay1}) that the centre
$r=0$ occurs at the null lines $x^\pm=0$, whereas according to the linearized
expression (\ref{lin}), there is a centre $r=0$ at \be B_-(\hat
x^+)^{-k}+B_+(-\hat x^-)^{-k}=-1/\epsilon \ee which, for $B_+<0$, $B_->0$, is a
spatial curve. This spatial centre may now be seen to be an artifact produced
by dropping the $\epsilon^2$ term in (\ref{lin}). Similar remarks apply to the
linearized trapping horizons $\partial_\pm r=0$, which do not appear if the
invariant definition $e^{-2\gamma}\partial_+ r\partial_- r = 0$ is used.

The problem may be regarded as due to the weak nature of the boundary
conditions imposed in \cite{hayward}. Imposing stronger boundary conditions
analogous to those in the perturbative analysis by Frolov \cite{fro} of the
related critical Roberts solution \cite{rob}, for instance that $\tilde r$ is
bounded as $\tau\to\pm\infty$, these spurious gauge modes may be excluded.

Note that this has not disproved the argument that the static solution lies at
the threshold between black holes and naked singularities. This still seems
likely, since a future-null singularity is just on the verge of being naked,
and a marginally future-trapped singularity is just on the verge of
disappearing inside a black hole. However, such solutions apparently cannot be
found by linear perturbations.

The conclusion is that the only genuine perturbative modes of the Hayward
solution are the $\omega_A$ modes in (\ref{tr}). As discussed in
\cite{hayward}, the imaginary part of the frequency ($\Im k \ge \sqrt{3}$) of
these modes produces echoing reminiscent of the numerical simulations of
Choptuik \cite{chop}.

\end{document}